\documentclass[twocolumn,pra,showpacs,superscriptaddress]{revtex4-1}
\usepackage{amssymb}
\usepackage{amsmath}
\usepackage{graphicx}
\usepackage{epsfig}

\begin{document}

\title{Transmission phase lapse in the non-Hermitian Aharonov--Bohm
interferometer near the spectral singularity}
\author{G. Zhang}
\affiliation{School of Physics, Nankai University, Tianjin 300071, China}
\author{X. Q. Li}
\affiliation{School of Physics, Nankai University, Tianjin 300071, China}
\author{X. Z. Zhang}
\affiliation{College of Physics and Materials Science, Tianjin Normal
University, Tianjin 300387, China}
\author{Z. Song}
\email{songtc@nankai.edu.cn}
\affiliation{School of Physics, Nankai University, Tianjin 300071, China}

\begin{abstract}
We study the effect of $\mathcal{PT}$-symmetric imaginary potentials
embedded in the two arms of an Aharonov-Bohm interferometer on the
transmission phase by finding an exact solution for a concrete tight-binding
system. It is observed that the spectral singularity always occurs at $k=\pm
\pi /2 $\ for a wide range of fluxes and imaginary potentials. Critical
behavior associated with the physics of the spectral singularity is also
investigated. It is demonstrated that the quasi-spectral singularity
corresponds to a transmission maximum and the transmission phase jumps
abruptly by $\pi $ when the system is swept through this point. Moreover, We
find that there exists a pulse-like phase lapse when the imaginary potential
approaches the boundary value of the spectral singularity.
\end{abstract}

\maketitle

%03.65.Nk, Scattering theory
%11.30.Er, Charge conjugation, parity, time reversal, and other discrete symmetries
%05.60.Gg, Quantum transport
%42.25.Bs, Wave propagation, transmission and absorption
%71.10.Fd, Lattice fermion models (Hubbard model, etc.)
%03.65.-w, Quantum mechanics
%72.20.Ee, Mobility edges; hopping transport
%03.65.Ge, Solutions of wave equations: bound states
%02.30.Fn, Several complex variables and analytic spaces
%42.25.Bs, Wave propagation, transmission and absorption
%73.43.-f£¬Quantum Hall effects
%73.40.Gk£¬Tunneling
%73.63.kv£¬Quantum dots
%73.23.Hk, Coulomb blockade; single-electron tunneling
%72.10.-d, Theory of electronic transport; scattering mechanisms
%73.23.-b, Electronic transport in mesoscopic systems
%85.35.Ds, Quantum interference devices

\section{Introduction}

\label{introduction}

Both the phase and the magnitude of a wavefunction are two important
quantities associated with quantum phenomena in nature. A direct application
is that the phase and magnitude of transmission can contain information
regarding the scattering center. For probability-based detection, we can
look back to the much earlier investigation of atomic structure, which led
to the development of the Rutherford model of the atom \cite{E. Rutherford11}
and eventually to the Bohr model. Now, the continued development of
technology makes it possible to experimentally investigate the transmission
phase, which contains information complementary to the transmission
probability \cite{A. Yacoby95,Yang Ji00,Yang Ji02,M. Sigrist04,M.
Avinun-Kalish05,M. Zaffalon08}. These measurements of the transmission phase
mainly focus on the so-called phase lapse phenomenon, which refers to an
abrupt\ jump in\ the transmission phase through a quantum dot between
transmission peaks \cite{G. Hackenbroich01}.

\begin{figure}[tbp]
\includegraphics[bb=30 300 550 800, width=0.43\textwidth, clip]{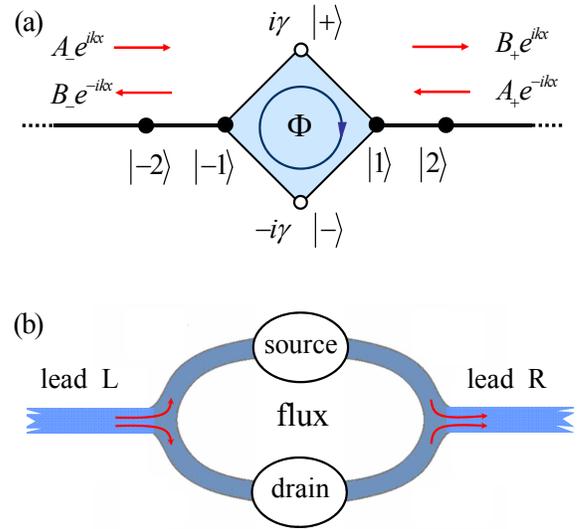}
\caption{(Color online) Schematic illustration of configuration of concerned
non-Hermitian AB interferometer. (a) It consists of a Hermitian
tight-binding square with an AB flux $\Phi$ and two semi-infinite chains as
the waveguides connecting to the scattering center. The non-Hermiticity of
the scattering center arises from the $\mathcal{PT}$ symmetric potentials $%
\pm i\protect\gamma $\ with respect to the axis along the leads. It is shown
that the transmission phase is sensitive to the flux when the system is near
the spectral singularity. (b) The model setup represents an open AB
interferometer with a source and drain embedded in the two arms, which can
be phenomenologically described by the type of tight-binding model in (a).
The flux breaks the balance between the source and drain and may result in
new transport behavior.}
\label{figure1}
\end{figure}

A non-Hermitian Hamiltonian can possess peculiar features that have no
counterpart in a closed Hermitian system. A typical example is
non-reciprocal dynamics, which has been observed in experiments \cite%
{Observe}. Especially, previous work \cite{LXQ} indicates that the
combination of magnetic field and non-Hermitian potential appears to have an
unexpected effect on particle transport behavior. The discovery of
non-Hermitian Hamiltonians with parity-time symmetry, which have a real
spectrum \cite{Bender}, has fundamentally boosted the research on the
complex extension of quantum mechanics \cite%
{Ann,JMP1,JPA1,JPA2,PRL1,JMP2,JMP3,JMP4,JPA3,JPA4,JPA5}. Recently, the
concept of spectral singularity of a non-Hermitian system has attracted
considerable attention \cite{PRA1,PRB1,Ali3,PRA3,JMP5,PRD1,PRA4,PRA5,PRA6},
motivated by the pioneering work of\ Mostafazadeh on the possible physical
relevance of the said concept \cite{PRL3}.\ Most previous works focus on
non-Hermitian systems with $\mathcal{PT}$-symmetry potentials \cite%
{PRA2,JPA6,Ali3,PRA13,prd2,prd3,prd4,prd5,prd6,prd7,prd8}, non-Hermitian
hopping amplitude \cite{PRA14,ZXZ,S. Longhi14}, and imaginary
particle-particle interaction strength \cite{LGR}.

In this study, we investigate the property of a non-Hermitian Aharonov-Bohm
(AB) interferometer with $\mathcal{PT}$-symmetric imaginary potentials
embedded in its two arms. We find that the spectral singularity with $k=\pm
\pi /2$\ exists for the system in a wide range of fluxes and imaginary
potentials. It is demonstrated that the quasi-spectral singularity
corresponds to\ a transmission maximum, and the transmission phase jumps
abruptly by $\pi $ when the system is swept through this point. Furthermore,
a pulse-like phase lapse exists when the imaginary potential approaches the
boundary value of the spectral singularity. This model can also suggest a
scheme for the realization of non-Hermitian imaginary hopping integral via
on-site imaginary potential. These findings can be exploited to detect
regions of criticality without having to undergo the spectral singularity
and to enhance interferometer sensitivity.

The remainder of this paper is organized as follows. In Section \ref{Model
and solutions}, we present the model setup and the solutions. In Section \ref%
{Spectral singularity}, the spectral singularity of the Hamiltonian is
examined. In Section \ref{Transmission phase lapse}, we study transmission
lapses near the spectral singularity. Finally, we present a summary and
discussion in Section \ref{Summary and discussion}.

\section{Model and solutions}

\label{Model and solutions}The non-Hermitian interferometer shown in Fig. %
\ref{figure1} is described by the Hamiltonian

\begin{eqnarray}
H &=&H_{\mathrm{0}}+H_{\mathrm{c}}  \label{H} \\
H_{\mathrm{0}} &=&\sum_{j=1}^{N}\left( \left\vert j\right\rangle
\left\langle j+1\right\vert +\left\vert -j\right\rangle \left\langle
-j-1\right\vert +\mathrm{H.c.}\right) , \\
H_{\mathrm{c}} &=&\frac{1}{\sqrt{2}}\sum_{\sigma =\pm }\left( e^{-i\sigma
\phi }\left\vert -1\right\rangle +e^{i\sigma \phi }\left\vert 1\right\rangle
\right) \left\langle \sigma \right\vert +\mathrm{H.c.}  \notag \\
&&+i\gamma \sum_{\sigma =\pm }\sigma \left\vert \sigma \right\rangle
\left\langle \sigma \right\vert ,
\end{eqnarray}%
which is a single-particle tight-binding model, where $\left\vert
j\right\rangle $\ denotes the site-state $j$. We consider the dimensionless
hopping integral $J=1$\ for simplicity. $H_{\mathrm{0}}$ represents the two\
leads, while $H_{\mathrm{c}}$\ is a non-Hermitian scattering center with an
AB flux $\Phi =4\phi $ enclosed by the two arms. The non-Hermiticity of the
scattering center arises from the $\mathcal{PT}$-symmetric potentials $\pm
i\gamma $\ with respect to the axis along the leads. This phenomenon can be
employed to phenomenologically depict an open interferometer, a
multi-terminal device \cite{G. Hackenbroich01}. For a tight-binding lattice
network, equivalence between the imaginary potential and the input (output)
lead is proposed \cite{L. Jin10,JL}. In another case, the imaginary
potential was added to an interferometer to introduce dephasing \cite{C.
Benjamin02}.

For the present model, we note that it has $\mathcal{PF}$-symmetry,%
\begin{equation}
\mathcal{\hat{P}\hat{F}}H\left( \mathcal{\hat{P}\hat{F}}\right) ^{-1}=H,
\end{equation}%
where the parity and flux flipping operators are

\begin{eqnarray}
\mathcal{\hat{P}} &:&\left\vert j\right\rangle \rightarrow \left\vert
-j\right\rangle , \\
\mathcal{\hat{F}} &:&\mathcal{\hat{F}}H\left( \phi \right) \mathcal{\hat{F}}%
^{-1}=H\left( -\phi \right) .
\end{eqnarray}%
For the flux-free case $\phi =0$, the system is $\mathcal{PT}$-symmetric
about the axis along the leads and $\mathcal{P}$-symmetric about the axis
through the locations of the imaginary potentials. A previous work \cite{L.
Jin12} shows that it is a probability-preserved system owing to balance
between $i\gamma $ and $-i\gamma $. It is presumable that the flux may break
such a balance and result in new transport behavior.

Based on the Bethe ansatz method, the solution of the Schrodinger equation
\begin{equation}
H\left\vert \psi _{k}\right\rangle =\varepsilon _{k}\left\vert \psi
_{k}\right\rangle  \label{S-eq}
\end{equation}%
takes the form%
\begin{equation}
\left\langle j\right. \left\vert \psi _{k}\right\rangle =\left\{
\begin{array}{cc}
A_{-}e^{ikj}+B_{-}e^{-ikj}, & \left( j\leqslant -1\right) \\
B_{+}e^{ikj}+A_{+}e^{-ikj}, & \left( j\geqslant 1\right)%
\end{array}%
\right. .
\end{equation}%
Eq. (\ref{S-eq}) results in $\varepsilon _{k}=2\cos k$ and two-component
spinor equation%
\begin{equation}
\chi \left(
\begin{array}{c}
B_{-} \\
B_{+}%
\end{array}%
\right) =\left\vert \chi \right\vert e^{i\theta \overrightarrow{n}\cdot
\overrightarrow{\sigma }}\left(
\begin{array}{c}
A_{-} \\
A_{+}%
\end{array}%
\right)  \label{rotation1}
\end{equation}%
where $\overrightarrow{n}=\left( n_{x},n_{y},0\right) $, $\overrightarrow{%
\sigma }=\left( \sigma _{x},\sigma _{y},\sigma _{z}\right) $ is a Pauli
matrix. When $\chi =0$, Eq. (\ref{rotation1}) is not useful. We will discuss
this later. Here the parameters are defined as
\begin{equation}
\left\{
\begin{array}{c}
\left\vert \chi \right\vert e^{i\theta }=\left\vert \eta \right\vert
^{2}-\xi ^{+}\xi ^{-}+i2\text{Im}\left( \eta \right) \sqrt{\xi ^{+}\xi ^{-}},
\\
n_{x}=\left( \xi ^{+}+\xi ^{-}\right) /\sqrt{4\xi ^{+}\xi ^{-}}, \\
n_{y}=i\left( \xi ^{+}-\xi ^{-}\right) /\sqrt{4\xi ^{+}\xi ^{-}},%
\end{array}%
\right. ,
\end{equation}%
and%
\begin{equation}
\left\{
\begin{array}{c}
\chi =\left[ \xi ^{+}\xi ^{-}-\left( \eta ^{\ast }\right) ^{2}\right]
e^{2ik}, \\
\eta =\left( e^{2ik}+1+\gamma ^{2}\right) e^{ik}, \\
\xi ^{\pm }=2\cos k\cos \left( 2\phi \right) \pm \gamma \sin \left( 2\phi
\right) ,%
\end{array}%
\right. .
\end{equation}

A straightforward implication of Eq. (\ref{rotation1}) is that it represents
the rotation operation of a two-component spinor. The direction of the
rotating axis $\overrightarrow{n}$ and angle $\theta $ could be complex. In
general, a given pair of arbitrary constants $A_{\pm }$ can generate a pair
of constants $B_{\pm }$, both of which together construct the eigenfunction $%
\left\vert \psi _{k}\right\rangle $. This indicates that the energy levels
are doubly degenerate. According to the theory of pseudo-Hermitian quantum
mechanics \cite{A. Mostafazadeh04}, a complete biorthogonal system requires
the construction of the eigenfunctions of $H^{\dag }$.

In parallel, we can perform the same procedure for the eigenfunction of the
Hamiltonian $H^{\dag }$. Similarly, we have the Schrodinger equation%
\begin{equation}
H^{\dag }\left\vert \overline{\psi }_{k}\right\rangle =\varepsilon
_{k}\left\vert \overline{\psi }_{k}\right\rangle  \label{S-eq*}
\end{equation}%
and the eigenfunction%
\begin{equation}
\left\langle j\right. \left\vert \overline{\psi }_{k}\right\rangle =\left\{
\begin{array}{cc}
\overline{A}_{-}e^{ikj}+\overline{B}_{-}e^{-ikj}, & \left( j\leqslant
-1\right) \\
\overline{B}_{+}e^{ikj}+\overline{A}_{+}e^{-ikj}, & \left( j\geqslant
1\right)%
\end{array}%
\right. .
\end{equation}%
The corresponding rotation equation of the two-component spinor reads%
\begin{equation}
\chi \left(
\begin{array}{c}
\overline{B}_{-} \\
\overline{B}_{+}%
\end{array}%
\right) =\left\vert \chi \right\vert e^{i\theta \overrightarrow{\overline{n}}%
\cdot \overrightarrow{\sigma }}\left(
\begin{array}{c}
\overline{A}_{-} \\
\overline{A}_{+}%
\end{array}%
\right) ,  \label{rotation2}
\end{equation}%
where only the unitary vector needs to be redefined as $\overrightarrow{%
\overline{n}}=\left( n_{x},-n_{y},0\right) $.

To construct the two degenerate eigenstates from a pair of arbitrary
constants $A_{\pm }$, it is beneficial to investigate the complete set of
two-component spinors. Taking the Hermitian conjugate of Eq. (\ref{rotation2}%
) and multiplying it by Eq. (\ref{rotation1}), we have%
\begin{eqnarray}
&&\left\vert \chi \right\vert ^{2}\left( \overline{B}_{-}^{\ast },\overline{B%
}_{+}^{\ast }\right) \left(
\begin{array}{c}
B_{-} \\
B_{+}%
\end{array}%
\right)  \notag \\
&=&\left( \overline{A}_{-}^{\ast },\overline{A}_{+}^{\ast }\right)
\left\vert \chi \right\vert ^{2}e^{-i\theta ^{\ast }\overrightarrow{%
\overline{n}}^{\ast }\cdot \overrightarrow{\sigma }}e^{i\theta
\overrightarrow{n}\cdot \overrightarrow{\sigma }}\left(
\begin{array}{c}
A_{-} \\
A_{+}%
\end{array}%
\right)  \notag \\
&=&\left\vert \chi \right\vert ^{2}\left( \overline{A}_{-}^{\ast },\overline{%
A}_{+}^{\ast }\right) \left(
\begin{array}{c}
A_{-} \\
A_{+}%
\end{array}%
\right) .
\end{eqnarray}%
It indicates that the orthonormal relationship between $\left( \overline{A}%
_{-},\overline{A}_{+}\right) $ and $\left( A_{-},A_{+}\right) $\ can be
transferred to that between $\left( \overline{B}_{-},\overline{B}_{+}\right)
$ and $\left( B_{-},B_{+}\right) $. This allows us to construct an entire
biorthogonal system based on an orthonormal set of two-component spinors.
Here, we povide an example by taking

\begin{equation}
\left(
\begin{array}{c}
\overline{A}_{-} \\
\overline{A}_{+}%
\end{array}%
\right) =\left(
\begin{array}{c}
A_{-} \\
A_{+}%
\end{array}%
\right) =\left(
\begin{array}{c}
\alpha _{-} \\
\alpha _{+}%
\end{array}%
\right) ,\text{ }\left(
\begin{array}{c}
-\alpha _{+}^{\ast } \\
\alpha _{-}^{\ast }%
\end{array}%
\right) ,
\end{equation}%
where $\left\vert \alpha _{+}\right\vert ^{2}+\left\vert \alpha
_{-}\right\vert ^{2}=1$. The corresponding spinors $\left( \overline{B}_{-},%
\overline{B}_{+}\right) $ and $\left( B_{-},B_{+}\right) $ are obtained
immediately. Then, for $\chi \neq 0$, we have two degenerate eigenfunctions
of $H$

\begin{equation}
\left\langle j\right. \left\vert \psi _{k}^{1}\right\rangle =\frac{C_{1}}{%
\sqrt{N}}\left\{
\begin{array}{cc}
\alpha _{-}e^{ikj}+\beta _{-}^{1}e^{-ikj}, & \left( j\leqslant -1\right) \\
\alpha _{+}e^{-ikj}+\beta _{+}^{1}e^{ikj}, & \left( j\geqslant 1\right)%
\end{array}%
\right. ,
\end{equation}%
and%
\begin{equation}
\left\langle j\right. \left\vert \psi _{k}^{2}\right\rangle =\frac{C_{2}}{%
\sqrt{N}}\left\{
\begin{array}{cc}
-\alpha _{+}^{\ast }e^{ikj}+\beta _{-}^{2}e^{-ikj}, & \left( j\leqslant
-1\right) \\
\alpha _{-}^{\ast }e^{-ikj}+\beta _{+}^{2}e^{ikj}, & \left( j\geqslant
1\right)%
\end{array}%
\right. ,
\end{equation}%
where $N$ is the system size. Here, the amplitudes%
\begin{eqnarray}
\beta _{\pm }^{1} &=&\frac{1}{\chi }\left[ \left( \left\vert \eta
\right\vert ^{2}-\xi ^{+}\xi ^{-}\right) \alpha _{\pm }+i2\text{Im}\left(
\eta \right) \xi ^{\mp }\alpha _{\mp }\right] , \\
\beta _{\pm }^{2} &=&\frac{1}{\chi }\left[ \pm \left( \left\vert \eta
\right\vert ^{2}-\xi ^{+}\xi ^{-}\right) \alpha _{\mp }^{\ast }\mp i2\text{Im%
}\left( \eta \right) \xi ^{\mp }\alpha _{\pm }^{\ast }\right] ,
\end{eqnarray}%
can be complex numbers and

\begin{equation}
C_{1}=\frac{\left\vert \chi \right\vert }{\sqrt{2\left\vert \chi \right\vert
^{2}+\Lambda _{++}}}\text{, }C_{2}=\frac{\left\vert \chi \right\vert }{\sqrt{%
2\left\vert \chi \right\vert ^{2}+\Lambda _{--}}},
\end{equation}%
are real numbers, where%
\begin{eqnarray}
&&\Lambda _{\sigma \sigma ^{\prime }}=4\text{Im}\left( \eta \right) \left(
\xi ^{+}-\xi ^{-}\right) [\text{Im}\left( \eta \right) \left( \xi
^{+}\left\vert \alpha _{\sigma ^{\prime }}\right\vert ^{2}-\xi
^{-}\left\vert \alpha _{-\sigma ^{\prime }}\right\vert ^{2}\right)  \notag \\
&&\sigma \left( \left\vert \eta \right\vert ^{2}-\xi ^{+}\xi ^{-}\right)
\text{Im}\left( \alpha _{-}\alpha _{+}^{\ast }\right) ],\left( \sigma
,\sigma ^{\prime }=\pm \right) .
\end{eqnarray}%
Accordingly, the eigenfunctions of $H^{\dag }$ can be expressed as%
\begin{equation}
\left\langle j\right. \left\vert \overline{\psi }_{k}^{1}\right\rangle =%
\frac{\overline{C}_{1}}{\sqrt{N}}\left\{
\begin{array}{cc}
\alpha _{-}e^{ikj}+\overline{\beta }_{-}^{1}e^{-ikj}, & \left( j\leqslant
-1\right) \\
\alpha _{+}e^{-ikj}+\overline{\beta }_{+}^{1}e^{ikj}, & \left( j\geqslant
1\right)%
\end{array}%
\right. ,
\end{equation}%
and%
\begin{equation}
\left\langle j\right. \left\vert \overline{\psi }_{k}^{2}\right\rangle =%
\frac{\overline{C}_{2}}{\sqrt{N}}\left\{
\begin{array}{cc}
-\alpha _{+}^{\ast }e^{ikj}+\overline{\beta }_{-}^{2}e^{-ikj}, & \left(
j\leqslant -1\right) \\
\alpha _{-}^{\ast }e^{-ikj}+\overline{\beta }_{+}^{2}e^{ikj}, & \left(
j\geqslant 1\right)%
\end{array}%
\right. ,
\end{equation}%
where $\overline{\beta }_{\pm }^{\lambda }\left( \phi \right) =\beta _{\pm
}^{\lambda }\left( -\phi \right) $, and real number $\overline{C}_{\lambda
}\left( \phi \right) =C_{\lambda }\left( -\phi \right) $, $\left( \lambda
=1,2\right) $, or

\begin{equation}
\overline{C}_{1}=\frac{\left\vert \chi \right\vert }{\sqrt{2\left\vert \chi
\right\vert ^{2}+\Lambda _{+-}}}\text{, }\overline{C}_{2}=\frac{\left\vert
\chi \right\vert }{\sqrt{2\left\vert \chi \right\vert ^{2}+\Lambda _{-+}}}%
\text{.}
\end{equation}

It is easy to check that $\left\langle \psi _{k}^{\lambda }\right.
\left\vert \psi _{k}^{\lambda }\right\rangle =1$. However, generally, $%
\left\langle \psi _{k}^{\lambda }\right. \left\vert \psi _{k}^{\lambda
^{\prime }}\right\rangle \neq 0$\ for $\lambda \neq \lambda ^{\prime }$,
unless the parameters are taken special value, e.g., $\gamma =0$ or $\sin
\left( 2\phi \right) =0$. In contrast, we have
\begin{equation}
\left\langle \overline{\psi }_{k}^{\lambda }\right. \left\vert \psi
_{k}^{\lambda ^{\prime }}\right\rangle =2\overline{C}_{\lambda }C_{\lambda
}\delta _{\lambda \lambda ^{\prime }}=\left\vert \chi \right\vert ^{2}%
\mathcal{G}\left( \left\vert \chi \right\vert \right) \delta _{\lambda
\lambda ^{\prime }},  \label{biorthogonal}
\end{equation}%
where one can see that $\mathcal{G}\left( \left\vert \chi \right\vert
\right) $\ is a nonzero bounded real function. It indicates that one can
always normalize the amplitudes $\overline{C}_{\lambda }$ and $C_{\lambda }$
to achieve a complete biorthogonal system.

Before we end this section, we would like to point out that: (i) It is not
helpful to choose eigenfunctions within each degeneracy subspace by using
the $\mathcal{PF}$-symmetry because $\mathcal{\hat{F}}$\ is not a Hermitian
operator. (ii) In the limit $\chi \rightarrow 0$, the biorthogonal
relationship in Eq. (\ref{biorthogonal}) tends to collapse, which implies
the emergence of the spectral singularity.

\section{Spectral singularity}

\label{Spectral singularity}

In this section, we will demonstrate the existence of spectral singularity
of the system and explore the feature of the solution at the critical point.
We start by considering the eigenfunctions of the system at the point $%
\left( k_{c},\phi _{c},\gamma _{c}\right) $ with $k_{c}=\pm \pi /2$ and
\begin{equation}
\sin ^{2}\left( 2\phi _{c}\right) =\gamma _{c}^{2},  \label{SS cond}
\end{equation}%
which lead to $\chi =0$. In this study, we only consider the non-Hermitian
case with $\gamma >0$. Thus, Eqs. (\ref{S-eq}) and (\ref{S-eq*}) can be
rewritten as%
\begin{eqnarray}
2\gamma _{c}^{4}\left( I\mp \sigma _{y}\right) \{\left(
\begin{array}{c}
A_{-} \\
A_{+}%
\end{array}%
\right) ,\left(
\begin{array}{c}
\overline{B}_{-} \\
\overline{B}_{+}%
\end{array}%
\right) \} &=&0,  \notag \\
2\gamma _{c}^{4}\left( I\pm \sigma _{y}\right) \{\left(
\begin{array}{c}
\overline{A}_{-} \\
\overline{A}_{+}%
\end{array}%
\right) ,\left(
\begin{array}{c}
B_{-} \\
B_{+}%
\end{array}%
\right) \} &=&0.  \label{solution of SS}
\end{eqnarray}%
The solutions of these equations are $A_{+}=\pm iA_{-}$, $\overline{A}%
_{+}=\mp i\overline{A}_{-}$, and $B_{+}=\mp iB_{-}$, $\overline{B}_{+}=\pm i%
\overline{B}_{-}$, which admit the eigenfunction of $H$ at the spectral
singularity%
\begin{equation}
\left\langle j\right. \left\vert \psi _{\pm \pi /2}^{c}\right\rangle =\frac{1%
}{\sqrt{2N}}\left\{
\begin{array}{cc}
e^{\pm i\frac{\pi }{2}j}, & \left( j\leqslant -1\right) \\
\pm ie^{\mp i\frac{\pi }{2}j}, & \left( j\geqslant 1\right)%
\end{array}%
\right. ,  \label{SS1}
\end{equation}%
where $\left\vert \psi _{\pm \pi /2}^{c}\right\rangle =\left\vert \psi _{\pm
\pi /2}^{1}\right\rangle =\left\vert \psi _{\mp \pi /2}^{2}\right\rangle $.
And the corresponding eigenfunction of $H^{\dagger }$ is%
\begin{equation}
\left\langle j\right. \left\vert \overline{\psi }_{\pm \pi
/2}^{c}\right\rangle =\frac{1}{\sqrt{2N}}\left\{
\begin{array}{cc}
e^{\pm i\frac{\pi }{2}j}, & \left( j\leqslant -1\right) \\
\mp ie^{\mp i\frac{\pi }{2}j}, & \left( j\geqslant 1\right)%
\end{array}%
\right. .  \label{SS2}
\end{equation}%
Similarly, $\left\vert \overline{\psi }_{\pm \pi /2}^{c}\right\rangle
=\left\vert \overline{\psi }_{\pm \pi /2}^{1}\right\rangle =\left\vert
\overline{\psi }_{\mp \pi /2}^{2}\right\rangle $. The physics of the
solutions\ is clear that $\left\vert \psi _{-\pi /2}^{c}\right\rangle $
describes self-sustained emission\ from the scattering center (lasering),
while $\left\vert \psi _{\pi /2}^{c}\right\rangle $ represents
reflectionless absorption of two incident plane waves (anti-lasering).

It can be readily checked whether

\begin{equation}
\left\langle \overline{\psi }_{\pm \pi /2}^{c}\right. \left\vert \psi _{\pm
\pi /2}^{c}\right\rangle =\left\langle \overline{\psi }_{\pm \pi
/2}^{c}\right. \left\vert \psi _{\mp \pi /2}^{c}\right\rangle =0,
\end{equation}%
which indicates that the complete biorthogonality\ of the eigenfunctions of $%
H$\ and $H^{\dag }$\ is destroyed at the points $\chi =0$. This so-called
spectral singularity has peculiar features for the present concrete system:
(i) The spectral singularity always occurs at the fixed $k$, which is
independent of the values of $\gamma $\ and $\phi $, whenever $\gamma $ is
within the range $\left[ -1,1\right] $. (ii) The transfer matrix $M$, which
is defined as
\begin{equation}
\left(
\begin{array}{c}
B_{+} \\
A_{+}%
\end{array}%
\right) =M\left(
\begin{array}{c}
A_{-} \\
B_{-}%
\end{array}%
\right)  \label{M}
\end{equation}%
has different property for the present scattering center. In fact, for Eq. (%
\ref{rotation1}) we have
\begin{equation}
\left( \eta -\eta ^{\ast }\right) \xi ^{+}\chi \left(
\begin{array}{c}
B_{+} \\
A_{+}%
\end{array}%
\right) =\widetilde{M}\left(
\begin{array}{c}
A_{-} \\
B_{-}%
\end{array}%
\right) ,
\end{equation}%
where the modified transfer matrix is%
\begin{equation}
\widetilde{M}=\left(
\begin{array}{cc}
-\left\vert \chi \right\vert ^{2} & \chi \left( \left\vert \eta \right\vert
^{2}-\xi ^{+}\xi ^{-}\right) \\
-\chi \left( \left\vert \eta \right\vert ^{2}-\xi ^{+}\xi ^{-}\right) & \chi
^{2}%
\end{array}%
\right) .
\end{equation}%
For $\left( \eta -\eta ^{\ast }\right) \xi ^{+}\chi \neq 0$, the determinant
of the transfer matrix is%
\begin{equation}
\det M=\frac{\xi ^{-}}{\xi ^{+}}=\frac{2\cos k\cos \left( 2\phi \right)
-\gamma \sin \left( 2\phi \right) }{2\cos k\cos \left( 2\phi \right) +\gamma
\sin \left( 2\phi \right) },
\end{equation}%
which is a function of $\left( k,\phi ,\gamma \right) $. We find that
whenever $k=\pm \pi /2$\ and $4\phi \neq 2n\pi $, we always have $\det M=-1$%
, which differs from the conclusion, $\det M=1$, in Refs. \cite{Ali1,Ali2}
for systems without flux. This implies that a scattering center subjected to
a magnetic field can have some special features. For the case of $\chi =0$,
which corresponds to the spectral singularity, we have%
\begin{equation}
M=\left(
\begin{array}{cc}
0 & -i \\
i & 0%
\end{array}%
\right) ,
\end{equation}%
from the solutions of Eq. (\ref{solution of SS}). This is in accordance with
the conclusion in \cite{Ali1,Ali2} that a signature of the spectral
singularity is $M_{22}=0$.

To exemplify the application of the present model, we will show that the AB
interferometer\ can be employed to realize a non-Hermitian imaginary hopping
integral in the tight-binding model. It has been reported that a
non-Hermitian center with imaginary hopping can be accessed by suitable
longitudinal modulations of gain/loss and propagation constants in
evanescently-coupled optical waveguide arrays, and it serves as a key
building block for realizing invisible defects in non-Hermitian
tight-binding lattices \cite{S. Longhi10}.

We begin with a simple case with $\phi =\pi /4$. Taking the following linear
transformation%
\begin{equation}
\left\{
\begin{array}{c}
\left\vert j\right\rangle =\underline{\left\vert j\right\rangle }\text{, }%
j\leqslant -1 \\
\left\vert +\right\rangle =\frac{1}{\sqrt{2}}e^{-i\pi /4}(\underline{%
\left\vert +\right\rangle }+\underline{\left\vert -\right\rangle })\text{,}
\\
\left\vert -\right\rangle =\frac{1}{\sqrt{2}}e^{i\pi /4}(\underline{%
\left\vert +\right\rangle }-\underline{\left\vert -\right\rangle })\text{,}
\\
\left\vert j\right\rangle =-i\underline{\left\vert j\right\rangle }\text{, }%
j\geqslant 1\text{,}%
\end{array}%
\right.  \label{transf1}
\end{equation}%
the original Hamiltonian (\ref{H}) can be written as%
\begin{eqnarray}
H_{\pi /4} &=&\sum_{j=1}^{N}(\underline{\left\vert j\right\rangle }%
\underline{\left\langle j+1\right\vert }+\underline{\left\vert
-j\right\rangle }\underline{\left\langle -j-1\right\vert })  \notag \\
&&+(\underline{\left\vert -1\right\rangle }\underline{\left\langle
+\right\vert }+\underline{\left\vert 1\right\rangle }\underline{\left\langle
-\right\vert })+\mathrm{H.c.}  \notag \\
&&+i\gamma (\underline{\left\vert +\right\rangle }\underline{\left\langle
-\right\vert }+\underline{\left\vert -\right\rangle }\underline{\left\langle
+\right\vert }),  \label{H EQ1}
\end{eqnarray}%
which reduces the AB ring to a non-Hermitian imaginary hopping dimer.
According to the above analysis, $H_{\pi /4}$ has a spectral singularity at $%
k=\pm \pi /2$ when $\gamma =1$.

\begin{figure}[tbp]
\includegraphics[bb=30 30 550 800, width=0.45\textwidth, clip]{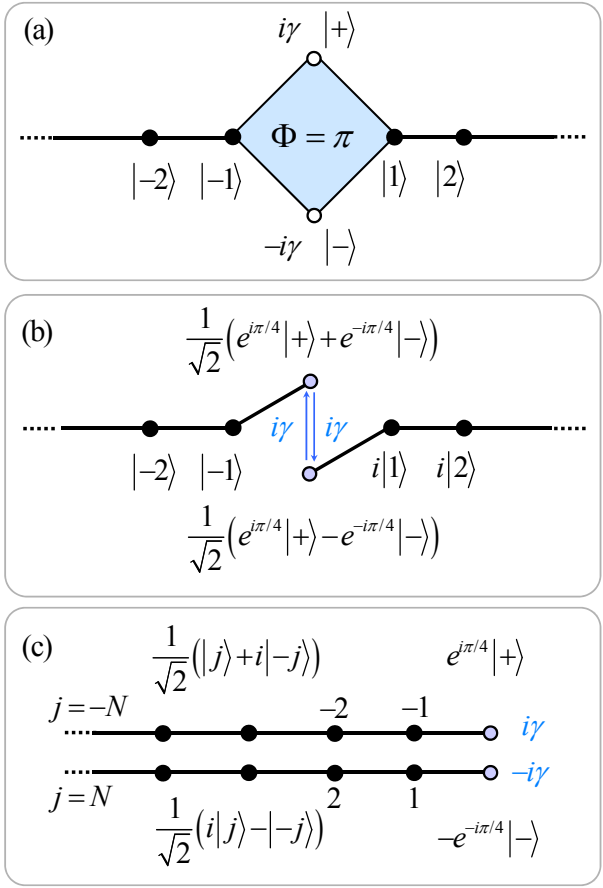}
\caption{(Color online) Schematic illustration of exemplified system.\ (a)
Non-Hermitian scattering center configuration with $\protect\phi =\protect%
\pi /4$, which consists of two on-site imaginary potentials $i\protect\gamma
$\ and $-i\protect\gamma $. (b) The equivalent Hamiltonian $H_{\protect\pi %
/4}$\ in Eq. (\protect\ref{H EQ1}), which is obtained via linear
transformation of Eq. (\protect\ref{transf1}). It represents a system with a
non-Hermitian imaginary hopping dimer, which has the hopping integral $i%
\protect\gamma $. (c) The equivalent Hamiltonian $H_{\protect\pi /4}$\ in
Eq. (\protect\ref{H EQ2}) is obtained via the two linear transformations of
Eqs. (\protect\ref{transf1}) and (\protect\ref{transf2}). It is shown that
the original Hamiltonian can be mapped to two separated Hamiltonians $H_{\pm
}$, describing semi-infinite chains with ending imaginary potentials $\pm i%
\protect\gamma $.}
\label{figure2}
\end{figure}

Taking the following linear transformation

\begin{equation}
\left\{
\begin{array}{c}
\underline{\left\vert j\right\rangle }=\frac{1}{\sqrt{2}}(\overline{%
\left\vert j\right\rangle }-\overline{\left\vert -j\right\rangle })\text{, }%
j\leqslant -1 \\
\underline{\left\vert +\right\rangle }=\frac{1}{\sqrt{2}}(\overline{%
\left\vert +\right\rangle }-\overline{\left\vert -\right\rangle })\text{,}
\\
\underline{\left\vert -\right\rangle }=\frac{1}{\sqrt{2}}(\overline{%
\left\vert +\right\rangle }+\overline{\left\vert -\right\rangle })\text{,}
\\
\underline{\left\vert j\right\rangle }=\frac{1}{\sqrt{2}}(\overline{%
\left\vert j\right\rangle }+\overline{\left\vert -j\right\rangle })\text{, }%
j\geqslant 1%
\end{array}%
\right.  \label{transf2}
\end{equation}%
the Hamiltonian $H_{\pi /4}$ is decomposed into two separate parts

\begin{eqnarray}
H_{\pi /4} &=&H_{+}+H_{-},  \label{H EQ2} \\
H_{+} &=&\sum_{j=-N}^{-1}\overline{\left\vert j\right\rangle }\overline{%
\left\langle j+1\right\vert }+\overline{\left\vert -1\right\rangle }%
\overline{\left\langle +\right\vert }+\mathrm{H.c.}  \notag \\
&&+i\gamma \overline{\left\vert +\right\rangle }\overline{\left\langle
+\right\vert }, \\
H_{-} &=&\sum_{j=1}^{N}\overline{\left\vert j\right\rangle }\overline{%
\left\langle j+1\right\vert }+\overline{\left\vert 1\right\rangle }\overline{%
\left\langle -\right\vert }+\mathrm{H.c.}-i\gamma \overline{\left\vert
-\right\rangle }\overline{\left\langle -\right\vert }.
\end{eqnarray}%
The physics of the models clearly describe semi-infinite chains with ending
imaginary potentials $\pm i\gamma $. Such systems have been studied
systematically in a previous work \cite{PRA14}, in which the result was the
solution given in Eq. (\ref{SS1}). Therefore, the dynamic behaviors,
self-sustained emission, and reflectionless absorption of wavepackets, can
emerge\ in the system $H_{\pi /4}$ as well.

\section{Transmission phase lapse}

\label{Transmission phase lapse}

In this section, we investigate another physical relevance of the spectral
singularity. We begin with the scattering problem of the AB interferometer,
which should shed some light on the dynamics of wavepackets in the critical
region. The eigenfunctions of the incident wave from left and right can be
obtained by taking

\begin{equation}
\left(
\begin{array}{c}
A_{-} \\
A_{+}%
\end{array}%
\right) =\left(
\begin{array}{c}
1 \\
0%
\end{array}%
\right) \text{, }\left(
\begin{array}{c}
0 \\
1%
\end{array}%
\right) .
\end{equation}%
We have two degenerate eigenfunctions of $H$

\begin{equation}
\left\langle j\right. \left\vert \psi _{k}^{\mathrm{L}}\right\rangle
=\left\{
\begin{array}{cc}
e^{ikj}+r_{\mathrm{L}}e^{-ikj}, & \left( j\leqslant -1\right) \\
t_{\mathrm{L}}e^{ikj}, & \left( j\geqslant 1\right)%
\end{array}%
\right. ,
\end{equation}%
and%
\begin{equation}
\left\langle j\right. \left\vert \psi _{k}^{\mathrm{R}}\right\rangle
=\left\{
\begin{array}{cc}
e^{-ikj}+r_{\mathrm{R}}e^{ikj}, & \left( j\geqslant 1\right) \\
t_{\mathrm{R}}e^{-ikj}, & \left( j\leqslant -1\right)%
\end{array}%
\right. .
\end{equation}%
The transmission and reflection\ amplitudes $t_{\mathrm{L,R}}$ and $r_{%
\mathrm{L,R}}$\ can be obtained from the corresponding $B_{\pm }$. These
amplitudes obey the relations%
\begin{equation}
r_{\mathrm{R}}\left( \phi \right) =r_{\mathrm{L}}\left( -\phi \right) =r_{%
\mathrm{L}}\text{, }t_{\mathrm{R}}\left( \phi \right) =t_{\mathrm{L}}\left(
-\phi \right) ,
\end{equation}

\begin{figure*}[tbph]
\begin{center}
\includegraphics[bb=90 260 500 570,width=0.4\textwidth,
clip]{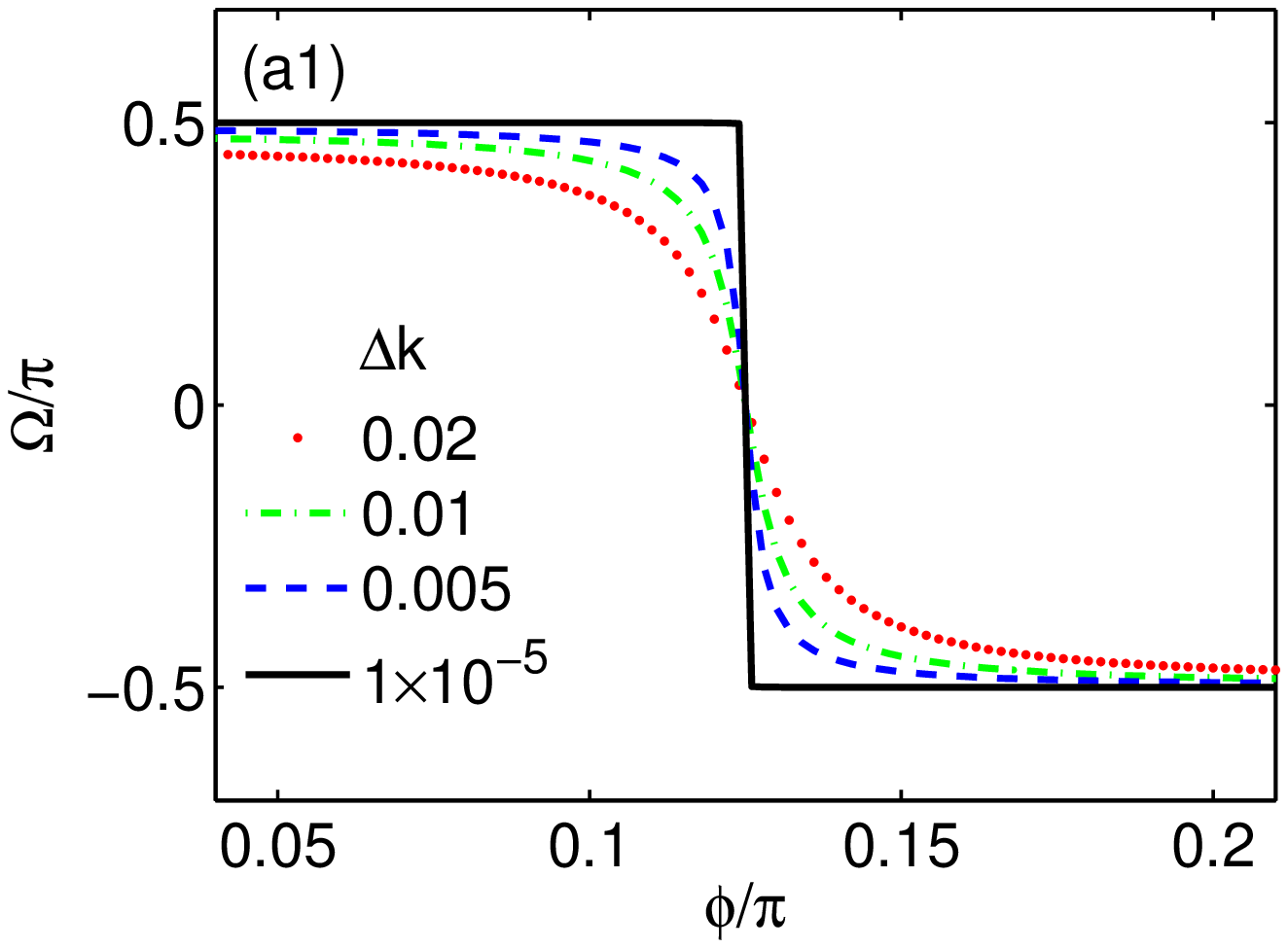}
\includegraphics[bb=90 260 500
570,width=0.4\textwidth,clip]{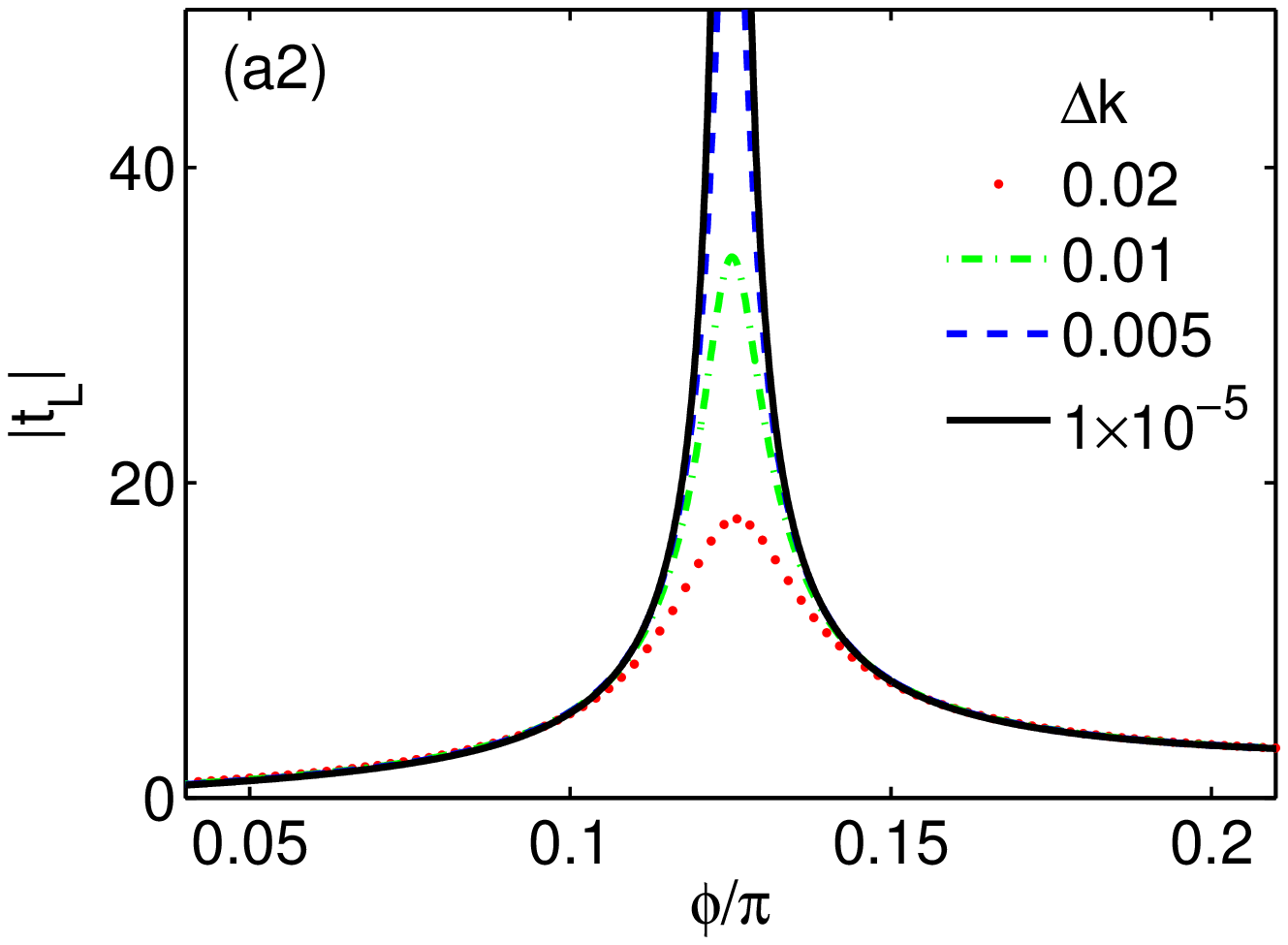}
\includegraphics[bb=90 260 500 570,width=0.4\textwidth,
clip]{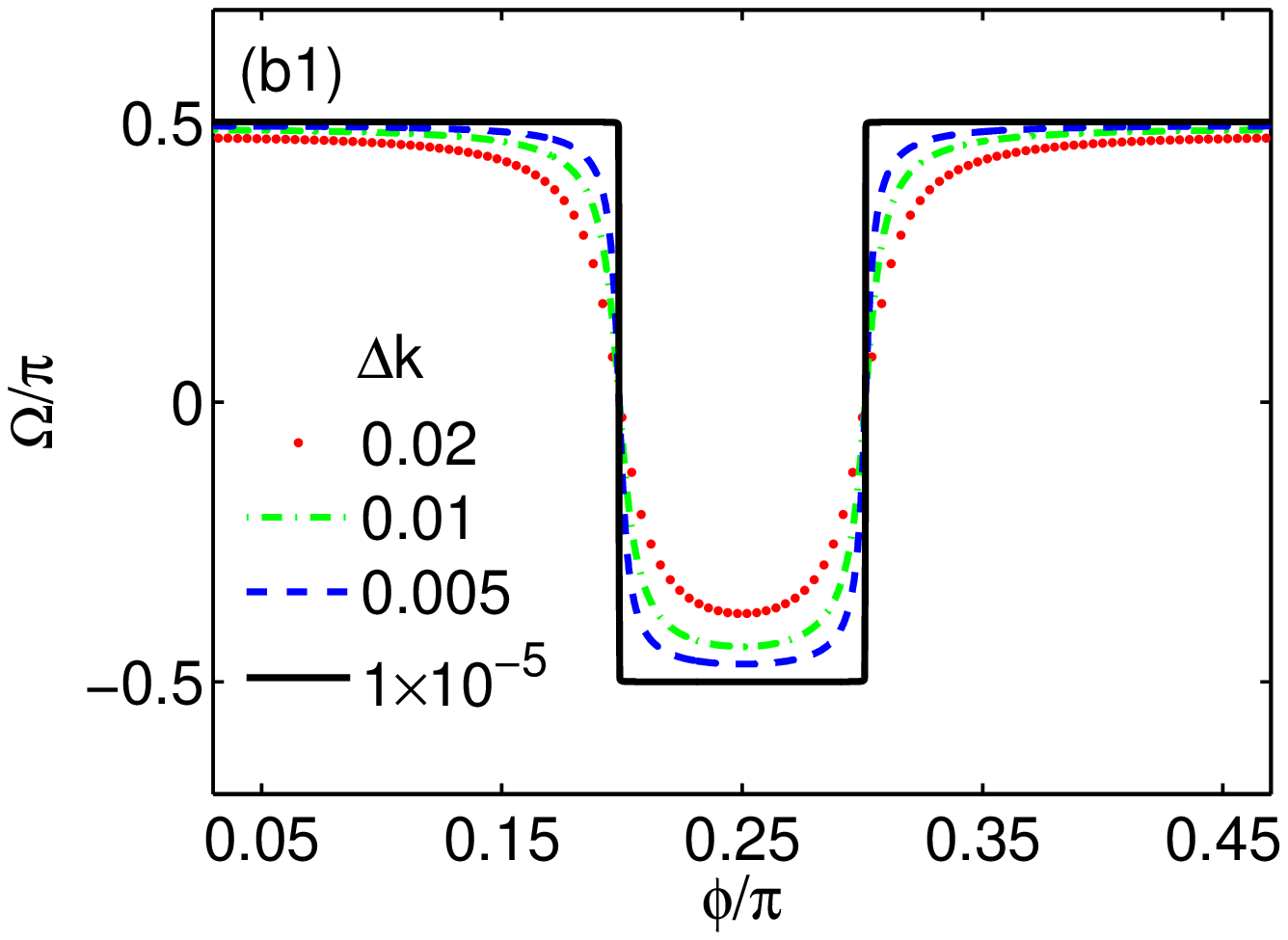}
\includegraphics[bb=90 260 500
570,width=0.4\textwidth,clip]{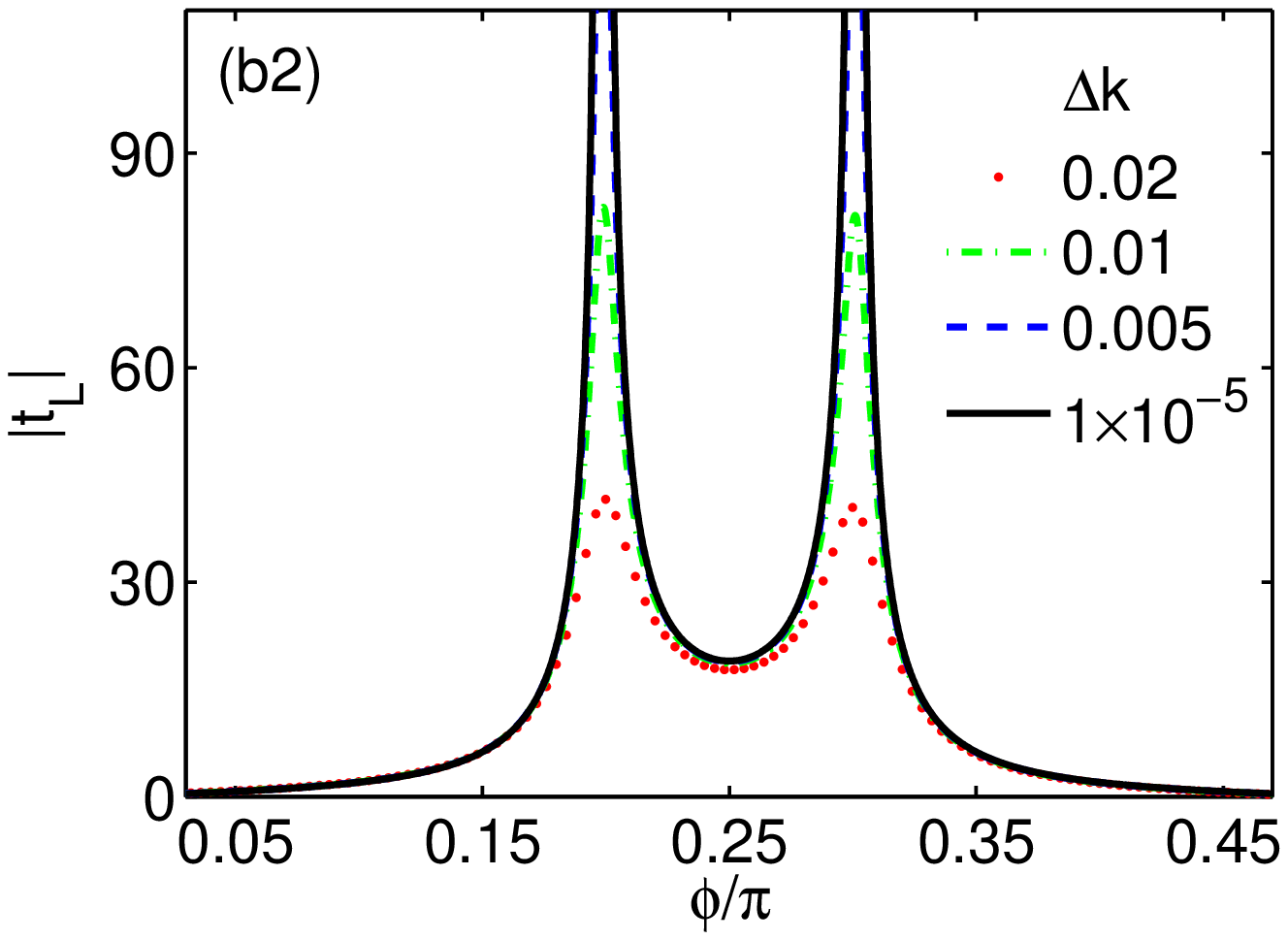}
\includegraphics[bb=90 260 500
570,width=0.4\textwidth,clip]{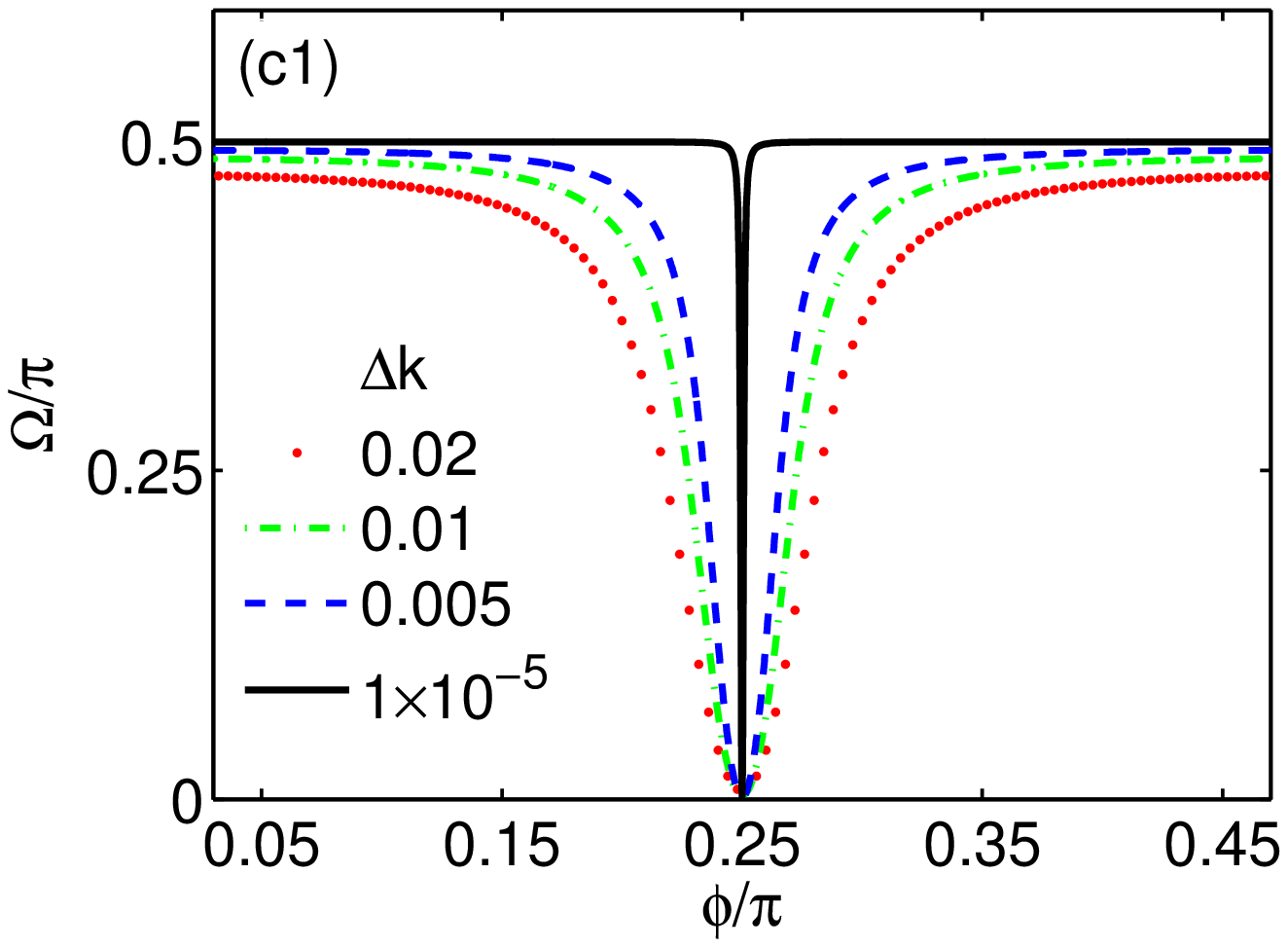}
\includegraphics[bb=90 260 500
570,width=0.4\textwidth,clip]{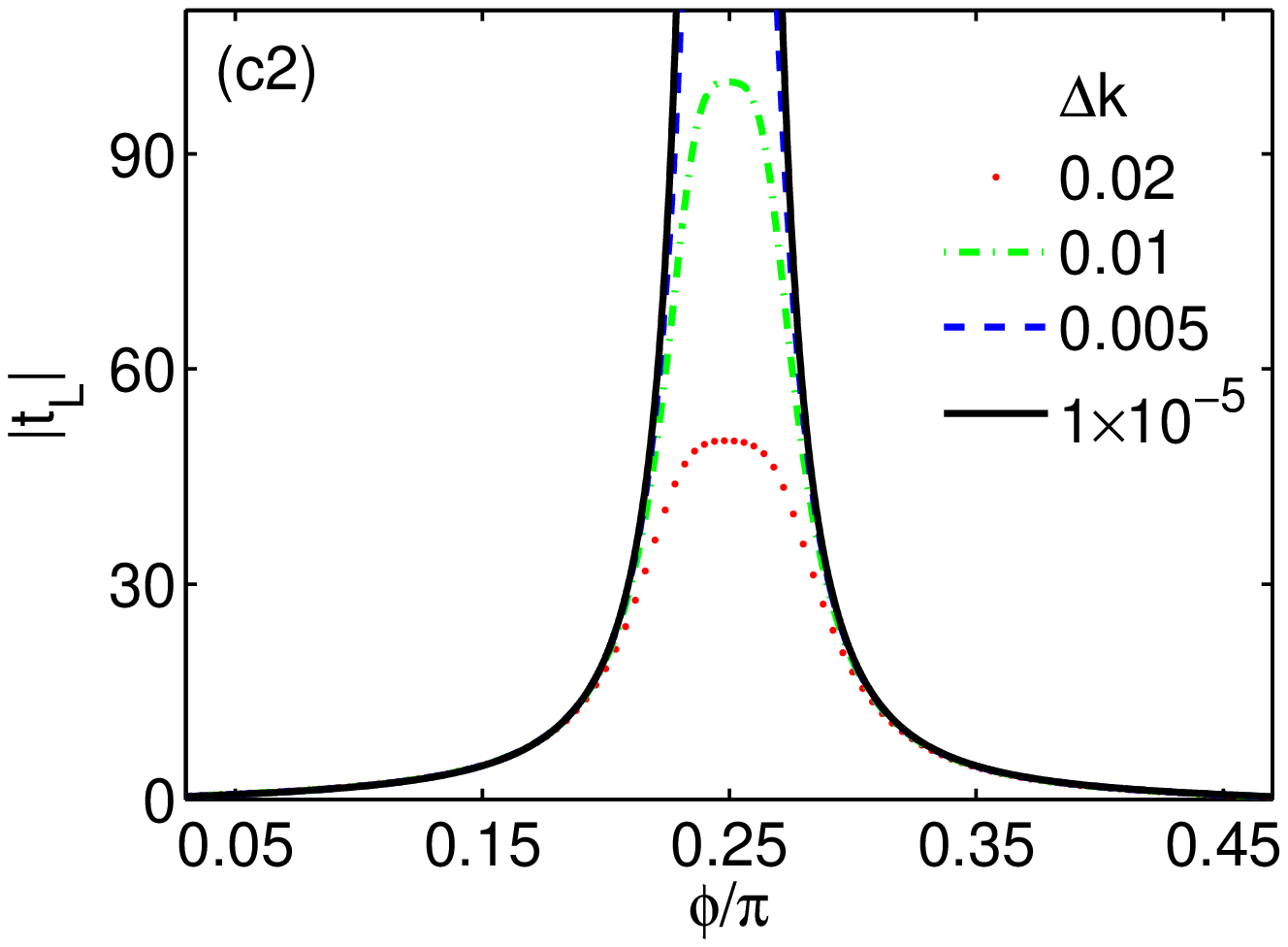}
\end{center}
\caption{(Color online) Plots of transmission amplitudes as functions of
flux near spectral singularity, which demonstrate two types of lapse of
transmission phase for various $k=k_{c}+\Delta k$. The phase and magnitude
of $t_{\mathrm{L}}$\ in Eq. (\protect\ref{t_L}) are plotted for (a) $\protect%
\gamma =0.707$, (b) $\protect\gamma =0.949$, (c) $\protect\gamma =1.00$. The
plot shows that the profiles of the transmission phase and the magnitude of $%
t_{\mathrm{L}}$\ are in agreement with our analysis based on the Eq. (%
\protect\ref{t_L_app}).}
\label{figure3}
\end{figure*}

\begin{figure}[tbp]
\includegraphics[bb=85 265 500 570, width=0.45\textwidth,
clip]{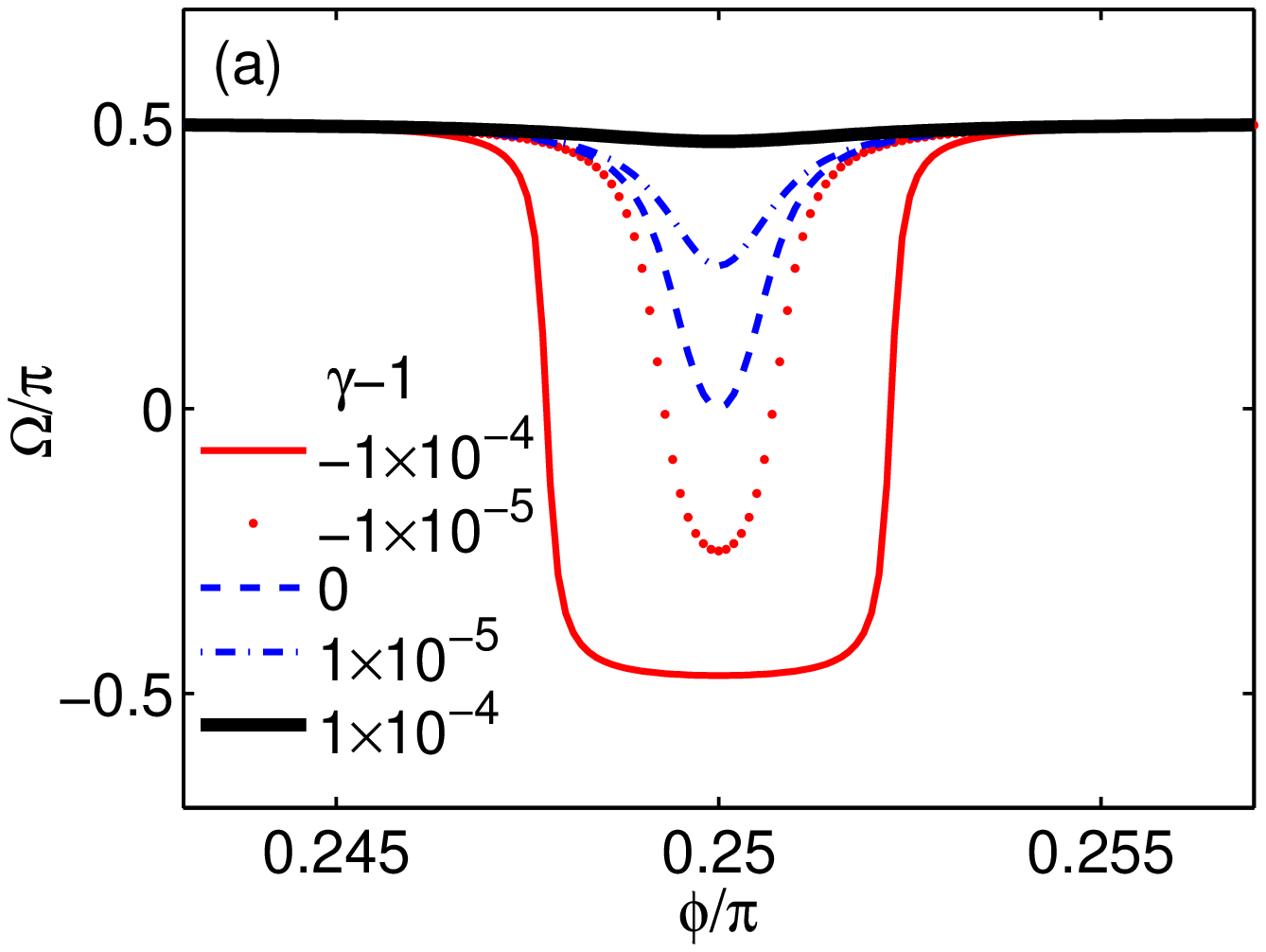}
\includegraphics[bb=85 265 500 570,
width=0.45\textwidth, clip]{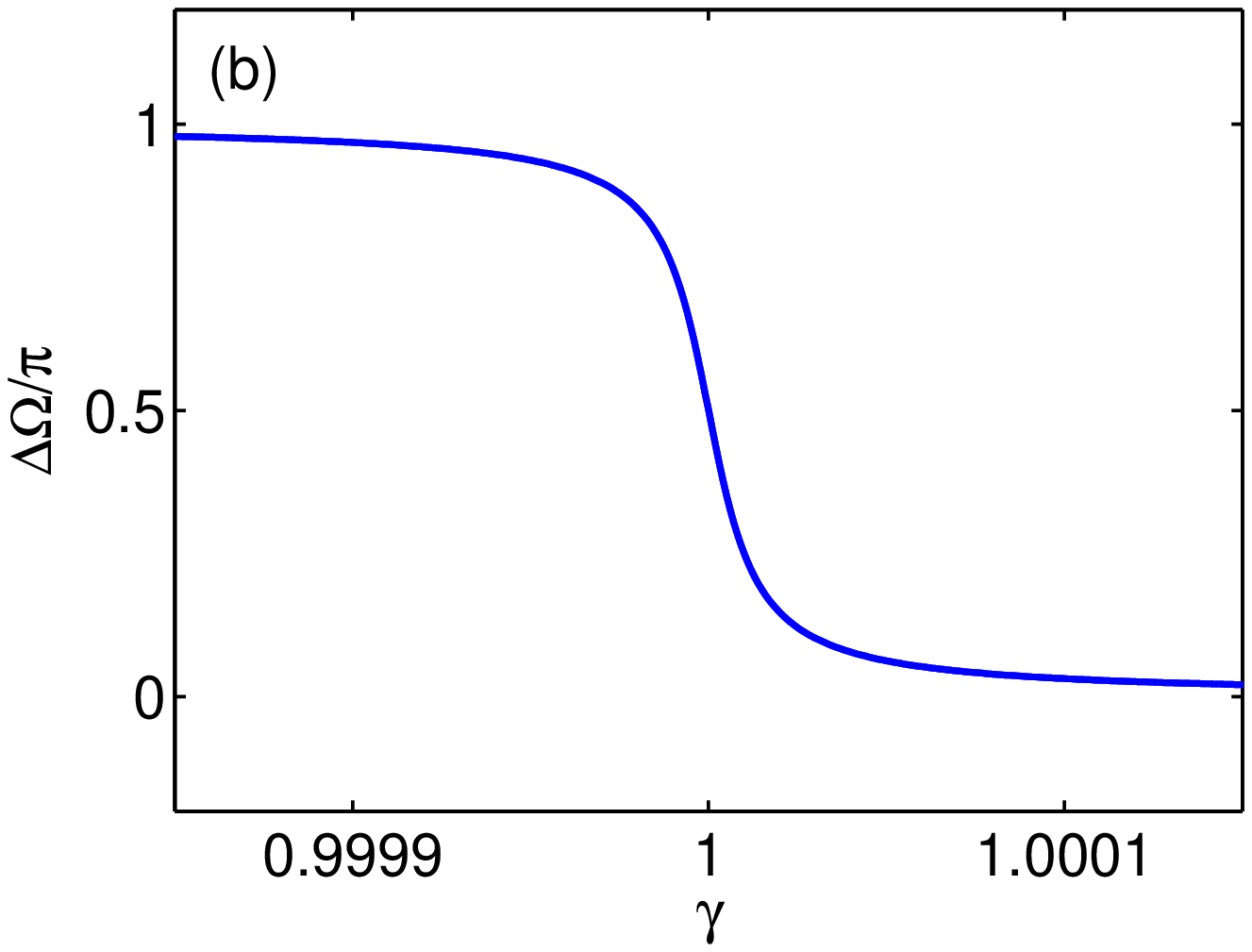}
\caption{(Color online) Crossover from $\protect\pi $\ lapse to zero lapse
for system with $\protect\gamma $\ around $1$.\ (a) Plots of transmission
phases for $k=\protect\pi /2+10^{-5}$ as functions of the flux near the
spectral singularity, which show the pulse-like lapse with different
heights. (b) The maximal phase shifts for $k=\protect\pi /2+10^{-5}$ as a
function of $\protect\gamma $, as obtained from $t_{\mathrm{L}}$\ in Eq. (%
\protect\ref{t_L}). The profiles of the transmission phase as functions of $%
\protect\phi $\ and $\protect\gamma $ are in agreement with our analysis
based on Eq. (\protect\ref{t_L_app}).}
\label{figure4}
\end{figure}
owing to $\mathcal{PF}$-symmetry and can be written in the explicit form
\begin{widetext}
\begin{equation}
r_{\mathrm{L}}=\frac{\left\vert e^{2ik}+1+\gamma ^{2}\right\vert ^{2}-\left[
4\cos ^{2}k\cos ^{2}\left( 2\phi \right) -\gamma ^{2}\sin ^{2}\left( 2\phi
\right) \right] }{e^{2ik}\left[ 4\cos ^{2}k\cos ^{2}\left( 2\phi \right)
-\gamma ^{2}\sin ^{2}\left( 2\phi \right) \right] -\left( e^{-2ik}+1+\gamma
^{2}\right) ^{2}},  \label{r_L}
\end{equation}%
and%
\begin{equation}
t_{\mathrm{L}}=\frac{\left[ e^{ik}\left( e^{2ik}+1+\gamma ^{2}\right) -%
\mathrm{C.c.}\right] \left( 2\cos k\cos \left( 2\phi \right) -\gamma \sin
\left( 2\phi \right) \right) }{e^{2ik}\left[ 4\cos ^{2}k\cos ^{2}\left(
2\phi \right) -\gamma ^{2}\sin ^{2}\left( 2\phi \right) \right] -\left(
e^{-2ik}+1+\gamma ^{2}\right) ^{2}}.  \label{t_L}
\end{equation}%
\end{widetext}In the vicinity of the spectral singularity $\left( k_{c},\phi
_{c},\gamma _{c}\right) $, we have

\begin{equation}
t_{\mathrm{L}}\approx \frac{\gamma _{c}^{2}}{\left\vert \rho \right\vert }%
\mathrm{sign}\left[ \gamma _{c}\sin \left( 2\phi _{c}\right) \right]
e^{i\Omega },  \label{t_L_app}
\end{equation}%
where%
\begin{equation}
\left\{
\begin{array}{c}
\rho =\sin \left( 4\phi _{c}\right) \left( \phi -\phi _{c}\right) +2\cos
\left( 4\phi _{c}\right) \left( \phi -\phi _{c}\right) ^{2} \\
-\left( \gamma -\gamma _{c}\right) +i\left( 2-\gamma _{c}^{2}\right) \left(
k-k_{c}\right) , \\
\Omega =\mathrm{Arg}\left( \rho \right) -\frac{\pi }{2},%
\end{array}%
\right. .  \label{ro}
\end{equation}%
The term $\left( \phi -\phi _{c}\right) ^{2}$\ is retained for\ the case of
very small $\sin \left( 4\phi _{c}\right) $. This approximate expression in
Eq. (\ref{t_L_app}) indicates that the transmission phase exhibits following
features.

(i) In the case of $0<\gamma <1$, there always exists spectral
singularities, for instance, at the point $\phi _{c}$ (or $\pi /2-\phi _{c}$)%
\textbf{,} $k_{c}=\pi /2$\ and $\gamma _{c}=\gamma $. We now consider the
transmission behavior of $k\sim k_{c}$, $\phi $\ varying in the vicinity of $%
\phi _{c}$. When $\gamma $\ is not close to $0$ and $1$ such that the term $%
\left( \phi -\phi _{c}\right) $\ is dominant in the real part of $\rho $,
the magnitude of $\rho $\ reaches a minimum, while its real part switches
its sign as $\phi $ passes the point $\phi _{c}$. According to Eq. (\ref%
{t_L_app}), these events lead to the magnitude of $t_{\mathrm{L}}$\ reaching
a maximum at $\phi =\phi _{c}$, while the phase $\Omega $\ jumps by $\pi $
in the case of $\left\vert \phi -\phi _{c}\right\vert \gg \left\vert \left(
2-\gamma _{c}^{2}\right) \left( k-k_{c}\right) /\sin \left( 4\phi
_{c}\right) \right\vert $. We can see that the phase shift becomes very
abrupt when $k$ is close to $k_{c}$. Then in the limit case, lapse of the
transmission phase is from $\pi /2$ to $-\pi /2$. Similarly, a lapse from $%
-\pi /2$ to $\pi /2$ should occur near the point $\pi /2-\phi _{c}$. This
implies two succeeding abrupt shifts when the two points $\phi _{c}$\ and $%
\pi /2-\phi _{c}$ are close to each other. Actually, when $\gamma $\ is
close to $1$, from Eq. (\ref{SS cond}), we have $\phi _{c}\approx \pi /4$.
According to Eq. (\ref{t_L_app}), the magnitude of $t_{\mathrm{L}}$\ reaches
a minimum at $\phi =n\pi /4$, but maxima at $\phi =\phi _{c}$, $\pi /2-\phi
_{c}$. The transmission phase can exhibit a pulse-like shift of height $\pi $%
,\ i.e., a lapse from $\pi /2$ to $-\pi /2$ to $\pi /2$. We will see from
the following analysis that as $\gamma _{c}$\ increases to $1$, the pulse
height decreases.

(ii) In the case of $\gamma >1$, there is no singularity. When $\gamma $\ is
not close to $1$ such that the term $\left( \gamma -1\right) $\ is dominant
in the real part of $\rho $, there is no lapse of the transmission phase as $%
\phi $ passes the point $\phi _{c}=\pi /4$. It is interesting to see what
happens to the crossover from (i) to (ii). To this end, we consider the
following case.

(iii) $\gamma =1$.\ For this case, we have $\phi _{c}=\pi /4$, the term of $%
\left( \phi -\phi _{c}\right) ^{2}$\ being dominant in the real part of $%
\rho $. The magnitude of $t_{\mathrm{L}}$\ reaches a maximum at $\phi =\pi
/4 $, while the transmission phase experiences two succeeding abrupt $\pi /2$
shifts as $\phi $\ varies, i.e., from $\pi /2$ to $0$ to $\pi /2$, similar
to a pulse of height $\pi /2$. This indicates crossover of the transmission
phase lapse from $\pi $\ to zero.

To demonstrate the above analysis, we plot the phase and magnitude from Eq. (%
\ref{t_L}) for several types of cases in Fig. \ref{figure3}. The figure
shows that our analysis is in accordance with the exact expression $t_{%
\mathrm{L}}$ when the system approaches the spectral singularity. Moreover,
we simulated the crossover from case (i) to (ii), as shown in Fig. \ref%
{figure4}. First, for various values of $\gamma $\ around $1$,\ we plot the
phase $\Omega \left( \phi \right) $\ as a function of $\phi $. Secondly, we
plot the maximal phase shift, which is defined as $\Delta \Omega =\mathrm{Max%
}\left[ \Omega \left( \phi \right) -\Omega \left( \pi /4\right) \right] $,
as a function of $\gamma $. The numerical results clearly show that the
transmission phase lapse is a good indicator of the transition between
systems with and without spectral singularity.\

\section{Summary and discussion}

\label{Summary and discussion}

In summary, we studied the non-Hermitian AB interferometer. On the basis of
the exact solution of a concrete tight-binding system, it is found that
there are fixed spectral singularities at $k=\pm \pi /2$ for a wide range of
fluxes and imaginary potentials. The critical behavior associated with the
physics of the spectral singularity exhibits two types of lapses of the
transmission phases, from $\pi /2$ to $-\pi /2$ and from $\pi /2$ to $-\pi
/2 $ to $\pi /2$.\ These phenomena can be exploited as a tool to detect the
regions of criticality without undergoing the spectral singularity and
enhance interferometer sensitivity. In addition, the concrete example also
suggested a scheme for realizing non-Hermitian imaginary hopping dimer with
the aid of on-site imaginary potential. This appears to imply that the
combination of $\mathcal{PT}$-symmetric non-Hermitian potential and magnetic
flux is crucial for such a phenomenon. Finally, this approach can be
extended to more generalized systems such as interferometers with longer
arms and complex potential as $\pm i\gamma \rightarrow \pm \left( V+i\gamma
\right) $, in which the spectral singularity should not be fixed.

\acknowledgments We acknowledge the support of the National Basic Research
Program (973 Program) of China under Grant No. 2012CB921900 and CNSF (Grant
No. 11374163).

\end{document}